\documentclass{elsarticle}  
\usepackage{hyperref}
\usepackage{xcolor}

\journal{Journal of New Astronomy}

\begin{document}

\begin{frontmatter}

\title{Investigation of pairing correlations on computed Gamow-Teller strength distributions and associated $\beta$-decay half-lives}


\author{Jameel-Un Nabi}
\address{University of Wah, Quaid Avenue, Wah Cantt 47040, Punjab, Pakistan.}

\author{Asim Ullah}
\address{Faculty of Engineering Sciences, Ghulam Ishaq Khan Institute of Engineering Sciences and Technology, Topi, 23640, KP, Pakistan.}
\cortext[mycorrespondingauthor]{Corresponding author}
\ead{asimullah844@mail.com}

\author{Muhammad Tahir}
\address{University of Wah, Quaid Avenue, Wah Cantt 47040, Punjab, Pakistan.}

\begin{abstract}
We investigate the effect of pairing correlations on the computed Gamow-Teller (GT) strength distributions and corresponding $\beta$-decay half-lives. The calculations are performed for a total of 47 $sd$-shell nuclei, for 20 $<$ A $<$ 30, employing the pn-QRPA model. Our calculations use three different values of pairing gaps computed using three different empirical formulae. The GT strength distribution and centroid values change considerably with change in the pairing gap values. This in turn leads to differences in computed half-lives. The pairing gaps computed using the mass dependent formula result in the calculated half-lives in better agreement with the measured data. 
\end{abstract}


\end{frontmatter}

\section{Introduction}
Core-collapse supernovae are considered to be one of the main contributors to the formation of heavier elements in the stellar matter. Weak interactions
processes, especially electron capture (\textit{ec}) and $\beta$-decay (\textit{bd}), are believed to control both lepton to baryon fraction ($Y_e$) and entropy of the stellar core and play a major role leading to the core-collapse of the massive stars \cite{bethe1990supernova,bethe1979equation}. The \textit{ec} effectively reduces the number of electrons that provide the pressure support, while \textit{bd} has a converse action. Both these processes have a direct influence on the overall $Y_e$ ratio of the core of the star. These weak interaction processes
are important, not only in the accurate determination of the structure of the
stellar core, but also play a vital role in the elemental abundance calculation and nucleosynthesis. The computation of these reactions under stelalr conditions largely depend on the accurate and reliable calculation of Gamow-Teller (GT) response of the nuclei \cite{bethe1979equation}.\\
The GT transitions are the most common spin–isospin ($\sigma$$\tau$) type weak interaction processes in atomic nuclei \cite{osterfeld1992nuclear}. In spherical coordinate system, the isospin has three components, $\tau_+$, $\tau_-$ and $\tau_0$. The $\tau_+$ corresponds to the GT transitions (GT$_+$) in $\beta^+$ or \textit{ec} direction, where conversion of proton to neutron takes place. The $\tau_-$ represents the GT$_-$ transition where a neutron is changed to a proton via $\beta$$^-$ decay. The component $\tau_0$ refers to inelastic neutrino-nucleus scattering for low neutrino energies.
At start of core-collapse, due to low densities ($\sim$ 10$^{10}$ g/cm$^3$) and temperatures (300 - 800 keV), the chemical potential of electrons and nuclear Q-value have comparable magnitudes. Under such scenario, the \textit{ec} rates are sensitive to the detailed description of the GT strength distributions. However at much higher core densities, once the chemical potential exceeds the Q-value, the \textit{ec} rates are governed by  the total GT strength and centroid values. Therefore, a detailed knowledge of the GT distributions is in order for a reliable computation of stellar weak rates and $\beta$-decay half-lives. \\  
The GT transitions can be extracted, experimentally, via different charge-exchange reactions e.g. (\textit{p}, \textit{n}) \cite{anderson1991gamow}, (\textit{n}, \textit{p}) \cite{el1994spin}, ($^3He$, \textit{t}) \cite{fujita1999mirror} and (\textit{t}, $^3He$) \cite{cole2006measurement}. The GT strength distributions of hundreds of unstable nuclei are required for modeling and simualtion of core-collapse supernovae. A nuclear model, preferably microscopic in nature, is required that can reliably  calculate GT strength distributions and associated $\beta$-deacy half-lifes in reasonable agreement  with the available experimental data. 
Several attempts have been made in the past  to investigate the $\beta$-decay properties. Few noticeable mentions are the computations based on gross theory \cite{takahashi1973beta}, QRPA e.g. (\cite{staudt1990second,hirsch1993microscopic,nabi1999microscopic,wang2016nuclear,marketin2007calculation,tan2017calculations}) and shell model (\cite{martinez1999shell}). The gross theory calculations are unable to provide details of nuclear structure of individual nuclei and adopt a statistical approach to determine the $\beta$-decay properties. The shell model and pn-QRPA are the two microscopic approaches, widely employed for the accurate and reliable computations of the $\beta$-decay half-lives and stellar weak rates. However, the former approach use the Brink-Axel hypothesis \cite{brink1958theory} to include the contribution of excited state GT strength distributions at high temperature. The pn-QRPA model can compute the exited state GT transitions without invoking the hypothesis.\\
Pairing of nucleons inside the nucleus is treated using the BCS approach. The value of the pairing gaps is one of the crucial parameters used in the pn-QRPA model \cite{staudt1990second,hirsch1993microscopic}. 
In this study, we investigate the effect of pairing gaps on the computed GT strength distributions and $\beta$-decay half-lives. We calculated the half-lives of 47 $sd$-nuclei with mass in the range 20 $<$ A $<$ 30, employing the pn-QRPA model. The pairing gaps were calculated using three different empirical formulae, to be discussed in the next section. The computed half-lives were then compared with experimental data \cite{audi2017nubase2016}.\\
The paper is structured as follows. Section~2 provides a brief overview of the formalism used in our calculation. Section~3 discusses our results. The final section contains summary and concluding remarks.

\section{Formalism}

We employed the proton-neutron quasiparticle random phase
approximation (pn-QRPA) model to study the  the impact of pairing gaps on the computed GT strength and associated $\beta$-decay half-lives. We computed the GT distributions and $\beta$-decay half-lives for a total of 47 $sd$-shell nuclei in the mass range A = 20-30, using the pn-QRPA model. A brief necessary formalism of the model is presented below:

The following Hamiltonian was employed to calculate the GT strength distributions:
\begin{equation} \label{H}
	H^{QRPA} = H^{sp} + V^{pair} + V^{pp}_{GT} + V^{ph}_{GT},
\end{equation}
where $H^{sp}$ stands for the single particle Hamiltonian and $V^{pair}$ is the nucleon-nucleon pairing interaction for which the BCS approximation, as discussed earlier,  was considered. $V_{GT}^{pp}$ and $V_{GT}^{ph}$ are the particle-particle (\textit{pp}) and particle-hole (\textit{ph}) GT interactions, respectively. The wave functions and energies of the single particle were treated in the Nilsson model \cite{nilsson1955binding} with the incorporation of the nuclear deformation. The oscillator constant was computed using $\hbar\omega=41A^{1/3}$. The Nilsson-potential parameters were chosen from Ref.~\cite{ragnarsson1984systematics}. $Q$-values were calculated from the mass excess values taken from the mass compilation of Audi et al. \cite{audi2017nubase2016}. The values of deformation parameter ($\beta_2$) were taken from Ref.~\cite{moller2016nuclear}.\\
We began with a spherical nucleon basis ($c^{\dagger}_{jm}$, $c_{jm}$), with $j$ as total
angular momentum with z-component $m$. 
The spherical basis was transformed to the (axial-symmetric) deformed basis, denoted by ($d^{\dagger}_{m\alpha}$, $d_{m\alpha}$), using the transformation equation
\begin{equation}\label{df}
	d^{\dagger}_{m\alpha}=\Sigma_{j}D^{m\alpha}_{j}c^{\dagger}_{jm},
\end{equation}
where $d^{\dagger}$ and $c^{\dagger}$ are particle creation operators in the deformed and spherical basis, respectively. The matrices $D^{m\alpha}_{j}$ were obtained by diagonalizing the Nilsson Hamiltonian. The BCS
computations for the proton and neutron systems was done separately. We took a constant pairing force of strength G ($G_p$ and $G_n$ for protons and neutrons, respectively),
\begin{eqnarray}\label{pr}
		V_{pair}=-G\sum_{jmj^{'}m^{'}}(-1)^{l+j-m}c^{\dagger}_{jm}c^{\dagger}_{j-m}\\ \nonumber
		(-1)^{l^{'}+j^{'}-m^{'}} c_{j^{'}-m^{'}}c_{j^{'}m^{'}},
\end{eqnarray}
the summation over $m$ and $m^{'}$ was limited to $m$, $m^{'}$ $>$ 0, and $l$ is the orbital angular momentum. A quasiparticle (q.p) basis $(a^{\dagger}_{m\alpha}, a_{m\alpha})$ was later introduced from the Bogoliubov transformation
\begin{equation}\label{qbas}
	a^{\dagger}_{m\alpha}=u_{m\alpha}d^{\dagger}_{m\alpha}-v_{m\alpha}d_{\bar{m}\alpha}
\end{equation}
\begin{equation}
	a^{\dagger}_{\bar{m}\alpha}=u_{m\alpha}d^{\dagger}_{\bar{m}\alpha}+v_{m\alpha}d_{m\alpha},
\end{equation}
where $\bar{m}$, $a^{\dagger}$ and $a$ represent the time reversed state of $m$, the q.p. creation and annihilation operator, respectively, which appear later in the RPA equation. The occupation amplitudes ($u_{m\alpha}$ and $v_{m\alpha}$) are computed using BCS approximation (satisfying $u^{2}_{m\alpha}$+$v^{2}_{m\alpha}$ = 1).\\
Within the pn-QRPA framework, the GT transitions are described in terms of phonon creation and one describes the QRPA phonons as
\begin{equation}\label{co}
	A^{\dagger}_{\omega}(\mu)=\sum_{pn}[X^{pn}_{\omega}(\mu)a^{\dagger}_{p}a^{\dagger}_{\overline{n}}-Y^{pn}_{\omega}(\mu)a_{n}a_{\overline{p}}],
\end{equation}
where the indices $n$ and $p$ stand for $m_{n}\alpha_{n}$ and $m_{p}\alpha_{p}$, respectively, and differentiating neutron and proton single-particle states. The summation was taken over all proton-neutron pairs satisfying $\mu=m_{p}-m_{n}$ and $\pi_{p}.\pi_{n}$=1, with $\pi$ representing parity. In Eq.~(\ref{co}), $X$ and $Y$ represent the forward- and backward-going amplitudes, respectively, and are the eigenfunctions of the  RPA matrix  equation.  In pn-QRPA theory, the proton-neutron residual interactions work through \textit{ph} and \textit{pp} channels, characterized by interaction constants $\chi$ and $\kappa$, respectively. The $ph$ GT force can be expressed as
\begin{equation}\label{ph}
	V^{ph}= +2\chi\sum^{1}_{\mu= -1}(-1)^{\mu}Y_{\mu}Y^{\dagger}_{-\mu},\\
\end{equation}
with
\begin{equation}\label{y}
	Y_{\mu}= \sum_{j_{p}m_{p}j_{n}m_{n}}<j_{p}m_{p}\mid
	t_- ~\sigma_{\mu}\mid
	j_{n}m_{n}>c^{\dagger}_{j_{p}m_{p}}c_{j_{n}m_{n}},
\end{equation}
and the $pp$ GT force as
\begin{equation}\label{pp}
	V^{pp}= -2\kappa\sum^{1}_{\mu=
		-1}(-1)^{\mu}P^{\dagger}_{\mu}P_{-\mu},
\end{equation}
with
\begin{eqnarray}\label{p}
	P^{\dagger}_{\mu}= \sum_{j_{p}m_{p}j_{n}m_{n}}<j_{n}m_{n}\mid
	(t_- \sigma_{\mu})^{\dagger}\mid
	j_{p}m_{p}>\times \nonumber\\
	(-1)^{l_{n}+j_{n}-m_{n}}c^{\dagger}_{j_{p}m_{p}}c^{\dagger}_{j_{n}-m_{n}},
\end{eqnarray}
where the remaining symbols have their usual meanings. 
Here, the different signs in \textit{ph} and \textit{pp} force reveal the opposite nature of these interactions i.e. \textit{pp} force is attractive while the \textit{ph} force is repulsive. The interaction constants $\chi$ and $\kappa$ were chosen in concordance with the suggestion given in Ref.~\cite{homma1996systematic}, following a $1/A^{0.7}$ relation. Authors in Ref.~\cite{homma1996systematic} performed a systematic study of the $\beta$-decay within the  framework of pn-QRPA and employed a schematic GT residual interaction. The \textit{ph} and \textit{pp} force were consistently included for both $\beta^+$ and $\beta^-$ directions, and their strengths were fixed as smooth functions of mass number A of nuclei in such a way that the calculation best reproduced the
observed $\beta$-decay properties of nuclei. Our computation further fulfilled the model independent Ikeda sum rule~\cite{ikeda1963s}. The reduced transition probabilities for GT transitions from the QRPA ground state
to one-phonon states in the daughter nucleus were obtained as
\begin{equation}
	B_{GT} (\omega) = |\langle \omega, \mu ||t_{+} \sigma_{\mu}||QRPA \rangle|^2.
\end{equation}
For further details and complete solution of Eq.~(\ref{H}), we refer to Hirsch et al. \cite{hirsch1993microscopic}. 
The partial $\beta$-decay half-lives were calculated using the following relation
\begin{eqnarray}
	t_{p(1/2)} =	\frac{D}{(g_A/g_V)^2f_A(Z, A, E)B_{GT}(E_j)+f_V(Z, A, E)B_F(E_j)},
\end{eqnarray}
where $E_j$ is the final state energy, $E$ = $Q$ - $E_j$, $g_A/g_V$ (= -1.254)\cite{warburton1994first} represents ratio of axial vector to the vector coupling constant and D = $\frac{2\pi^3 \hbar^7 ln2}{g^2_V m^5_ec^4} = 6295 s$. Phase space factors $f_A(Z, A, E)$ ($f_V(Z, A, E)$) are the Fermi integral for axial vector (vector) transitions. $B_{GT}$ and $B_F$ are the reduced transition probabilities for the GT and Fermi transitions, respectively.
The $\beta$-decay half-lives were finally determined by summing up all transition probabilities to states in the daughter nucleus with excitation energies lying within the $Q$ window:
\begin{equation}
	T_{1/2} = \left(\sum_{0 \le E_j \le Q} \frac{1}{t_{p(1/2)}}\right)^{-1}.
\end{equation} 
To investigate the impact of pairing correlations on the calculated GT strengths and half-lives, we have used three different empirical formulae for computing the pairing gaps. The first one is the mass dependent formula, which is same for both proton and neutron, given as $\triangle_{nn}=\triangle_{pp}={12/\sqrt A}$ MeV~\cite{hardy2009superallowed}. The second formula consists of three terms, computes different pairing gaps for protons and neutrons and is given in terms of neutron and proton separation energies as:
\begin{equation}
	\bigtriangleup_{pp} =\frac{1}{4}(-1)^{Z+1}[S_p(A+1, Z+1)-2S_p(A, Z)+S_p(A-1, Z-1)]
\end{equation}
\begin{equation}
	\bigtriangleup_{nn} =\frac{1}{4}(-1)^{A-Z+1}[S_n(A+1, Z) - 2S_n(A, Z) + S_n(A-1, Z)]
\end{equation}
The third prescription is a systematic five term formula as a function of binding energies, given as;
\begin{eqnarray}
	\bigtriangleup_{nn} = \frac{1}{8}[B(N-2, Z) - 4B(N-1, Z) + 6B(N, Z) - 4B(N+1, Z) + B(N+2, Z)]
\end{eqnarray}
\begin{eqnarray}
	\bigtriangleup_{pp} = \frac{1}{8}[B(Z-2, N) - 4B(Z-1, N) + 6B(Z, N) - 4B(Z+1, N) + B(Z+2, N)]
\end{eqnarray}
The binding energies were taken from the recent mass compilation of Ref.~\cite{wang2021ame}. We refer to the first, second and third formula for computation of pairing gaps as TF, 3TF and 5TF, respectively, here onwards. 
\section{Results and Discussion}
In this work, we used the pn-QRPA model to investigate the impact of pairing gaps on the computed GT strength and $\beta$-decay half-lives. We performed the calculations for a total of 47 $sd$-shell nuclei with masses in the range 20 $<$ A $<$ 30, employing the three different schemes of pairing gaps denoted by TF, 3TF and 5TF, as discussed in the previous section. Out of  these 47 cases, 32 decay by electron emission ($\beta^-$) while 15 are unstable to positron emission ($\beta^+$ decay). We also compare our computed half-lives with the experimental data taken from Ref.~\cite{audi2017nubase2016}, referred to as Aud17 here onwards.\\
Fig.~1 shows the computed GT strength distributions for the three selected $\beta^-$ cases, $^{29}$Al, $^{27}$Ne and $^{22}$F, employing the three different schemes of pairing gaps. It can be seen that the strength distributions change considerably once the value of the pairing gap is changed. For the case $^{29}$Al, the strength distributions using TF and 3TF formulae are more or less the same. However, the strength changes considerably for 5TF scheme. The strength for the cases $^{27}$Ne and $^{22}$F do changes for all the three schemes. Fig.~2 depicts similar results for the $\beta^+$ cases $^{23}$Al, $^{20}$Na and $^{21}$Mg.
It is to be noted that only the strength within the Q$_{{\beta^-}/{\beta^+}}$ window has been shown is these two figures. \\
Table~1 displays the computed pairing gaps and half-lives using the three different formulae for the $\beta^-$ cases, whereas Table~2 shows the corresponding results for the $\beta^+$ cases. The computed half-lives are compared with the experimental data of Aud17, shown in the last column of these tables. It can be seen that the computed half-lives using the three formulae are in decent comparison for cases such as $^{20}$O, $^{24,26,29}$Ne and $^{28,30}$Mg. However, significant difference in the computed half-lives were noted for some other cases e.g. $^{23,26}$F, $^{27}$Ne, $^{25}$Na and $^{27}$P. In comparison with Aud17, the TF computed half-lives are in overall better agreement. The calculated half-lives using TF scheme were in comparison with Aud17, within 100\% deviation for 29 cases, within 500\% for 13 cases and the rest of the cases fall within 1000\% deviation. Similarly, the computed half-lives using 3TF (5TF) scheme were in agreement with Aud17, within 100\% for 24 (24) cases, within 500\% for 8 (7) cases, 1 (3) cases lies within 1000\% and the rest were above 1000\% deviation. Our study suggests that the computed half-lives are altered by the choice of pairing gaps. The reasons behind the differences in computed half-lives with different pairing gap values may be traced to the GT strength distributions and centroid values which we discuss next.\\
Table~3 and Table~4 display the values of total GT strength and energy centroid of the computed GT strength distributions employing the three schemes of pairing gaps for the $\beta^-$ and $\beta^+$ decays, respectively. The cut-off energy in daughter is also shown, in the last column of these two tables. The cut-off value lies within $Q_{\beta}$ window (see column 2 of Table 3 \& 4). It can be seen that changing the value of pairing gap slightly alter the total strength for most of the cases. However, at times, the total strength is significantly increased/decreased with change in the $\bigtriangleup_{nn}$($\bigtriangleup_{pp}$) values. For example, the total strength for $^{27}$Ne computed using the 5TF scheme is significantly smaller than those computed using TF and 3TF schemes. Similarly, the total strength for $^{22,23}$F using the TF scheme is considerably bigger than the one calculated using 3TF scheme. Similar difference were found in the centroid values. The smaller total strength value of $^{27}$Ne using 5TF leads to smaller weak rate and correspondingly higher computed half life as compare to the other two schemes (see Table~1). In the same way, the higher strength values of $^{22,23}$F using TF leads to smaller half-lives. The re-distribution of GT strength is another factor which effect the computed centroids and half life values. At times the GT strength shifts to higher excitation energy than cut-off value shown in the last column of Table 3 \& 4. For $^{25}$Na, the 3TF scheme places the centroid value at considerably higher excitation energy than TF and 5TF schemes. This leads to very smaller rate and correspondingly very high computed half life (Table~1). Similar trend were found for the other cases as well.\\   
Table~5 depicts the state by state computed GT strength, branching ratio (I) and partial half life ($t^{par}_{1/2}$) values for the $\beta^+$ decay of $^{27}$P. The branching ratio was calculated using:
\begin{eqnarray}
	I = \frac{T_{1/2}}t^{par}_{1/2} \times 100 (\%)
\end{eqnarray}
where $T_{1/2}$ represents the total half life. In case of TF and 5TF schemes, the high branching ratios at lower excitation energies leads to lower centroid values of 8.13 and 7.66, respectively as compared to 8.82 of 3TF scheme. These lower centroid values result in higher rates and correspondingly smaller half-lives. Moreover, the 3TF scheme computes very low total strength value which leads to higher calculated half life.\\
It is worth mentioning that the empty places in all the tables (Table~1-5) are because of two reasons: either we were unable to compute the pairing gaps due to missing mass excess values in Aud17 or the GT strength were lying outside the Q$_{{\beta^-}/{\beta^+}}$ window.

\begin{table}[]\label{T1}
	\scriptsize \caption{Calculated pairing gaps and $\beta$$^-$ decay half-lives, also compared with the measured data \cite{audi2017nubase2016}.  }
	
	\begin{tabular}{c|ccccc|cccc}
		\hline
		&&\multicolumn{4}{c}{Pairing Gaps (MeV)}&\multicolumn{4}{c}{Half-lives (s)}\\
		\hline
		Nuclei&$\bigtriangleup^{TF}_{nn=pp}$&$\bigtriangleup^{3TF}_{nn}$&$\bigtriangleup^{3TF}_{pp}$&$\bigtriangleup^{5TF}_{nn}$&$\bigtriangleup^{5TF}_{pp}$&$T^{TF}_{\textcolor{red}{1/2}}$&$T^{3TF}_{\textcolor{red}{1/2}}$&$T^{5TF}_{\textcolor{red}{1/2}}$&$T^{Aud17}_{\textcolor{red}{1/2}}$ \\
		\hline
		$^{20}$O  & 2.68 & 1.86 & 2.80 & 1.79 & 2.95 & 3.39E+01 & 3.13E+01 & 2.88E+01 & 1.35E+01 \\
		$^{21}$O  & 2.62 & 1.71 & 2.88 & 1.75 & 2.83 & 3.94E+00 &  ---        &  ---        & 3.42E+00 \\
		$^{22}$O  & 2.56 & 1.79 & 3.41 & 1.60 & 3.35 & 6.68E+00 & 4.82E+01 & 7.72E+00 & 2.25E+00 \\
		$^{23}$O  & 2.50 & 1.40 & 3.33 & 1.51 & 3.22 & 1.17E-01 & 1.42E+00 & 1.22E-01 & 9.70E-02 \\
		$^{24}$O  & 2.45 & 1.60 & 3.10 & 1.61 & 3.40 & 9.44E-02 & 4.61E-01 & 1.10E-01 & 7.74E-02 \\
		$^{21}$F  & 2.62 & 1.09 & 3.09 & 1.20 & 2.87 & 1.42E+00 & 8.08E-01 &  ---        & 4.16E+00 \\
		$^{22}$F  & 2.56 & 1.30 & 2.78 & 1.42 & 2.31 & 7.22E+00 & 1.70E+01 & 5.73E+01 & 4.23E+00 \\
		$^{23}$F  & 2.50 & 1.52 & 3.30 & 1.29 & 2.78 & 8.66E-01 & 3.60E+03 & 4.33E+03 & 2.23E+00 \\
		$^{24}$F  & 2.45 & 1.05 & 3.14 & 1.02 & 2.46 & 2.68E+00 & 2.05E+01 & 5.91E+00 & 3.84E-01 \\
		$^{26}$F  & 2.35 & 1.01 & 3.46 & 0.89 & 2.56 & 4.62E-02 & 2.76E+00 & 1.47E+00 & 8.20E-03 \\
		$^{27}$F  & 2.31 & 0.50 & 4.26 & 0.76 & 3.18 & 4.28E-02 & 2.38E-01 & 6.86E-02 & 4.90E-03 \\
		$^{29}$F  & 2.23 &---   &---      &   ---   &  ---    & 2.34E-03 &    ---      &   ---       & 2.50E-03 \\
		$^{23}$Ne & 2.50 & 2.21 & 1.84 & 2.15 & 1.69 & 1.64E+01 & 2.10E+01 & 2.87E+01 & 3.71E+01 \\
		$^{24}$Ne & 2.45 & 2.10 & 2.27 & 1.81 & 2.29 & 2.22E+02 & 2.24E+02 & 2.24E+02 & 2.03E+02 \\
		$^{25}$Ne & 2.40 & 1.53 & 1.81 & 1.45 & 1.86 & 6.95E-01 & 3.50E-01 & 3.48E-01 & 6.02E-01 \\
		$^{26}$Ne & 2.35 & 1.36 & 2.13 & 1.48 & 2.12 & 5.05E-01 & 5.05E-01 & 5.21E-01 & 1.97E-01 \\
		$^{27}$Ne & 2.31 & 1.59 & 1.62 & 1.45 & 1.46 & 8.41E-02 & 1.36E-01 & 4.25E+00 & 3.15E-02 \\
		$^{28}$Ne & 2.27 & 1.29 & 2.62 & 1.28 & 2.20 & 1.52E-02 & 1.89E-02 & 1.50E-02 & 2.00E-02 \\
		$^{29}$Ne & 2.23 & 1.27 & 2.46 & 1.29 & 1.94 & 7.81E-03 & 7.65E-03 & 8.03E-03 & 1.47E-02 \\
		$^{30}$Ne & 2.19 & 1.31 & 2.69 & 1.31 & 2.34 & 5.61E-03 & 4.17E-03 & 5.62E-03 & 7.22E-03 \\
		$^{25}$Na & 2.40 & 1.37 & 2.32 & 1.26 & 2.33 & 1.22E+01 & 5.53E+06 & 1.43E+01 & 5.91E+01 \\
		$^{26}$Na & 2.35 & 1.15 & 1.91 & 1.12 & 2.00 & 3.26E+00 & 1.15E+00 & 1.19E+00 & 1.07E+00 \\
		$^{27}$Na & 2.31 & 1.09 & 2.09 & 1.05 & 2.27 & 2.56E-01 & 5.82E-01 & 3.30E-01 & 3.01E-01 \\
		$^{28}$Na & 2.27 & 1.01 & 1.28 & 0.88 & 1.39 & 1.06E-01 & 3.39E-02 & 2.18E-01 & 3.05E-02 \\
		$^{29}$Na & 2.23 & 0.75 & 2.12 & 0.89 & 2.07 & 4.72E-02 & 2.99E-02 & 3.46E-02 & 4.41E-02 \\
		$^{30}$Na & 2.19 & 1.04 & 1.77 & 1.10 & 1.50 & 5.24E-02 & 1.82E-02 & 1.19E-02 & 4.84E-02 \\
		$^{27}$Mg & 2.31 & 1.68 & 2.09 & 1.70 & 2.08 & 1.87E+02 &    ---      &     ---     & 5.66E+02 \\
		$^{28}$Mg & 2.27 & 1.73 & 2.45 & 1.80 & 2.40 & 6.07E+03 & 4.03E+03 & 6.43E+03 & 7.53E+04 \\
		$^{29}$Mg & 2.23 & 1.89 & 1.48 & 1.78 & 1.52 & 1.14E+00 & 1.43E+00 & 1.86E+00 & 1.30E+00 \\
		$^{30}$Mg & 2.19 & 1.68 & 2.11 & 1.77 & 2.12 & 5.25E-01 & 5.26E-01 & 5.26E-01 & 3.13E-01 \\
		$^{29}$Al & 2.23 & 1.35 & 2.34 & 1.32 & 2.33 & 2.01E+02 & 2.01E+02 & 1.65E+01 & 3.94E+02 \\
		$^{30}$Al & 2.19 & 1.28 & 1.54 & 1.19 & 1.72 & 8.98E+00 & 2.67E+00 & 5.45E+00 & 3.62E+00 \\
		\hline
	\end{tabular}
\end{table}

\begin{table}[]
	\scriptsize \caption{Same as Table~1 but for $\beta$$^+$ decay cases.}
	\begin{tabular}{c|ccccc|cccc}
		\hline
		&&\multicolumn{4}{c}{Pairing Gaps (MeV)}&\multicolumn{4}{c}{Half-lives (s)}\\
		\hline
		Nuclei&$\bigtriangleup^{TF}_{nn=pp}$&$\bigtriangleup^{3TF}_{nn}$&$\bigtriangleup^{3TF}_{pp}$&$\bigtriangleup^{5TF}_{nn}$&$\bigtriangleup^{5TF}_{pp}$&$T^{TF}_{\textcolor{red}{1/2}}$&$T^{3TF}_{\textcolor{red}{1/2}}$&$T^{5TF}_{\textcolor{red}{1/2}}$&$T^{Aud17}_{\textcolor{red}{1/2}}$ \\
		\hline
		$^{20}$Na & 2.68 & 2.25 & 1.32 & 2.25 & 1.19 & 1.04E+00 & 4.39E-01 & 5.19E-01 & 4.48E-01 \\
		$^{21}$Na & 2.62 & 2.25 & 3.37 & 2.05 & 2.74 & 1.31E+02 & 1.08E+02 & 2.09E+02 & 2.24E+01 \\
		$^{20}$Mg & 2.68 & ---     &   ---   &    ---  &    ---  & 7.57E-01 &      ---    &      ---   & 9.30E-02 \\
		$^{21}$Mg & 2.62 & 3.13 & 1.07 & 2.94 & 1.13 & 1.30E-01 & 2.49E-01 & 2.45E-01 & 1.19E-01 \\
		$^{23}$Al & 2.50 & 1.83 & 2.13 & 1.74 & 2.14 & 6.34E-01 & 2.36E-01 & 1.82E-01 & 4.70E-01 \\
		$^{24}$Al & 2.45 & 1.68 & 1.82 & 1.80 & 1.51 & 8.29E-01 & 5.35E-01 & 8.91E-01 & 2.05E+00 \\
		$^{24}$Si & 2.45 & 2.34 & 2.04 & 2.51 & 2.00 & 1.49E-01 & 8.93E-02 & 1.46E-01 & 1.40E-01 \\
		$^{25}$Si & 2.40 & 2.52 & 1.20 & 2.51 & 1.09 & 5.94E-02 &     ---     &   ---       & 2.20E-01 \\
		$^{26}$Si & 2.35 & 2.44 & 1.97 & 2.42 & 1.80 & 1.68E+01 & 1.62E+01 & 1.60E+01 & 2.25E+00 \\
		$^{27}$Si & 2.31 & 2.40 & 1.64 & 2.77 & 1.65 & 1.17E+01 &    ---      &     ---     & 4.15E+00 \\
		$^{26}$P  & 2.35 & 1.93 & 1.09 & 1.82 & 1.10 & 6.30E-02 & 3.26E-02 & 1.77E-02 & 4.37E-02 \\
		$^{27}$P  & 2.31 & 2.05 & 1.57 & 2.00 & 1.68 & 6.81E-01 & 1.06E+01 & 9.51E-01 & 2.60E-01 \\
		$^{28}$P  & 2.27 & 2.16 & 1.67 & 2.29 & 1.44 & 5.54E-01 & 3.64E-01 & 6.47E-01 & 2.70E-01 \\
		$^{28}$S  & 2.27 & 2.16 & 1.48 & 2.44 & 1.77 & 1.26E-01 & 1.12E-01 & 1.17E-01 & 1.25E-01 \\
		$^{30}$S  & 2.19 & 2.40 & 1.44 & 2.22 & 1.51 & 7.27E+00 & 5.35E+00 & 5.51E+00 & 1.18E+00\\
		\hline
	\end{tabular}
\end{table}

\begin{table}[]\label{T3}
	\scriptsize \caption{Computed total GT strength and centroid values for $\beta$$^-$ decays employing the three pairing gaps (TF, 3TF and 5TF). The Q-values and cut-off energy values in daughter are shown in the second and last columns, respectively. }
		\centering
	\begin{tabular}{c|c|ccc|ccc|c}
		\hline
	& 	&\multicolumn{3}{c}{Total Strength (arb. units)}&\multicolumn{3}{c}{Centroids (MeV)}& {Cut-off Energy} \\
		\hline
		Nuclei&Q$_{\beta^-}$&$\sum$ GT$^{TF}$ &$\sum$ GT$^{3TF}$ &$\sum$ GT$^{5TF}$ &$\bar{E}$$^{TF}$&$\bar{E}$$^{3TF}$&$\bar{E}$$^{5TF}$&E$_x$ (MeV) \\
		\hline
		$^{20}$O& 3.81 & 0.13 & 0.14 & 0.15  & 0.50  & 0.49  & 0.49  & 0.58  \\
		$^{21}$O& 8.11 & 0.06 &   ---   &   ---    & 2.90  &    ---   &   ---    & 4.07  \\
		$^{22}$O& 6.49 & 2.32 & 2.45 & 2.85  & 5.04  & 5.34  & 5.53  & 6.39  \\
		$^{23}$O& 11.33 & 0.99 & 1.81 & 0.99  & 6.21  & 8.20  & 6.22  & 11.09 \\
		$^{24}$O& 10.96 & 3.32 & 3.15 & 2.98  & 5.64  & 7.24  & 5.68  & 10.33 \\
		$^{21}$F& 5.68 & 2.12 & 2.22 &    ---   & 2.72  & 2.04  &     ---  & 4.18  \\
		$^{22}$F& 10.82 & 0.66 & 0.11 & 0.28  & 7.75  & 7.41  & 9.04  & 10.63 \\
		$^{23}$F& 8.44 & 0.68 & 0.08 & 0.09  & 4.14  & 7.22  & 7.31  & 8.33  \\
		$^{24}$F& 13.49 & 0.08 & 0.36 & 0.64  & 7.94  & 11.77 & 11.38 & 13.39 \\
		$^{26}$F& 18.17 & 1.20 & 0.54 & 2.10 & 11.82 & 15.99 & 16.59 & 18.01 \\
		$^{27}$F& 18.40 & 4.70 & 1.34 & 5.64 & 14.87 & 13.77 & 16.33 & 18.40 \\
		$^{29}$F& 21.75 & 9.39 & ---     &    ---   & 15.85 &   ---    &   ---    & 21.40 \\
		$^{23}$Ne& 4.38 & 0.20 & 0.18 & 0.14  & 1.96  & 2.24  & 2.32  & 3.29  \\
		$^{24}$Ne& 2.47 & 0.68 & 0.69 & 0.69  & 1.32  & 1.33  & 1.32  & 2.21  \\
		$^{25}$Ne& 7.32 & 1.08 & 0.88 & 0.88  & 3.84  & 3.63  & 3.61  & 7.28  \\
		$^{26}$Ne& 7.34 & 1.94 & 1.97 & 1.98  & 3.21  & 3.24  & 3.23  & 6.99  \\
		$^{27}$Ne& 12.57 & 2.38 & 2.09 & 0.04  & 8.12  & 8.01  & 6.67  & 12.49 \\
		$^{28}$Ne& 12.29 & 5.22 & 3.40 & 5.58  & 5.91  & 6.19  & 5.99  & 12.29 \\
		$^{29}$Ne& 15.72 & 0.77 & 0.77 & 0.75  & 3.17  & 3.12  & 3.24  & 5.54  \\
		$^{30}$Ne& 14.81 & 8.67 & 4.18 & 8.63  & 9.24  & 6.05  & 9.21  & 14.79 \\
		$^{25}$Na& 3.83 & 0.41 & 0.13 & 0.36  & 1.02  & 3.70  & 1.19  & 3.70  \\
		$^{26}$Na& 9.35 & 0.47 & 1.76 & 1.74  & 5.23  & 6.49  & 6.62  & 9.13  \\
		$^{27}$Na& 9.07 & 1.43 & 0.47 & 1.54  & 6.42  & 4.13  & 6.99  & 9.05  \\
		$^{28}$Na& 14.03 & 1.06 & 1.09 & 1.49  & 6.93  & 4.99  & 9.80  & 13.63 \\
		$^{29}$Na& 13.28 & 1.34 & 3.09 & 3.03  & 9.76  & 10.41 & 10.34 & 13.28 \\
		$^{30}$Na& 17.36 & 0.53 & 0.77 & 3.04  & 8.80  & 7.29  & 11.06 & 16.38 \\
		$^{27}$Mg& 2.61 & 0.43 &   ---   &   ---    & 1.03  &  ---     &   ---    & 1.08  \\
		$^{28}$Mg& 1.83 & 0.11 & 0.28 & 0.10  & 1.15  & 1.40  & 1.16  & 1.76  \\
		$^{29}$Mg& 7.60 & 0.25 & 0.31 & 0.29  & 2.75  & 3.59  & 4.70  & 7.52  \\
		$^{30}$Mg& 6.98 & 1.61 & 1.63 & 1.63  & 2.44  & 2.46  & 2.45  & 5.46  \\
		$^{29}$Al& 3.69 & 0.25 & 0.26 & 0.27  & 2.35  & 2.36  & 1.05  & 3.51  \\
		$^{30}$Al& 8.57 & 0.50 & 0.49 & 0.48  & 5.91  & 4.89  & 5.81  & 8.18 \\
		\hline
	\end{tabular}
\end{table}

\begin{table}[]
	\scriptsize  \caption{Same as Table~3 but for $\beta$$^+$ decay cases.}
	\begin{tabular}{c|c|ccc|ccc|c}
		\hline
	&	&\multicolumn{3}{c}{Total Strength (arb. units)}&\multicolumn{3}{c}{Centroids (MeV)}& {Cut-off Energy} \\
		\hline
		Nuclei&Q$_{\beta^+}$&$\sum$ GT$^{TF}$ &$\sum$ GT$^{3TF}$ &$\sum$ GT$^{5TF}$ &$\bar{E}$$^{TF}$&$\bar{E}$$^{3TF}$&$\bar{E}$$^{5TF}$&E$_x$ (MeV) \\
		\hline
		$^{20}$Na&13.89 & 1.71 & 1.47 & 1.83  & 9.17  & 8.17 & 9.26  & 13.71 \\
		$^{21}$Na&3.55 & 0.20 & 0.37 & 0.20  & 0.54  & 0.80 & 1.33  & 3.44  \\
		$^{20}$Mg&10.63 & 1.84 &  ---    &   ---    & 6.48  &  ---    &    ---   & 10.47 \\
		$^{21}$Mg&13.09 & 2.46 & 2.29 & 2.30  & 6.59  & 7.48 & 7.46  & 12.68 \\
		$^{23}$Al&12.22 & 1.91 & 2.39 & 2.05  & 10.06 & 9.35 & 8.82  & 12.21 \\
		$^{24}$Al&13.88 & 2.16 & 2.03 & 0.27  & 8.93  & 8.44 & 7.27  & 13.39 \\
		$^{24}$Si&10.79 & 6.76 & 6.66 & 7.12  & 7.28  & 6.50 & 7.42  & 10.73 \\
		$^{25}$Si&12.74 & 4.50 &  ---    &    ---   & 9.33  &  ---    &   ---    & 12.66 \\
		$^{26}$Si&5.07 & 2.11 & 2.57 & 2.65  & 3.29  & 3.30 & 3.34  & 4.45  \\
		$^{27}$Si&4.81 & 0.71 &  ---    &   ---    & 0.58  &  ---   &  ---     & 4.70  \\
		$^{26}$P &18.11 & 2.99 & 0.90 & 2.97 & 15.23 & 6.88 & 12.53 & 17.82 \\
		$^{27}$P &11.33 & 2.07 & 0.24 & 2.39  & 8.13  & 8.82 & 7.66  & 11.28 \\
		$^{28}$P &14.35 & 1.26 & 1.21 & 1.64  & 8.05  & 7.49 & 10.21 & 13.96 \\
		$^{28}$S &11.22 & 8.05 & 8.82 & 8.55  & 7.83  & 8.10 & 8.00  & 11.15 \\
		$^{30}$S &6.14 & 1.66 & 2.37 & 2.35  & 3.66  & 4.17 & 4.16  & 6.11 \\
		\hline
	\end{tabular}
\end{table}

\begin{table}[]
	\scriptsize\caption{The state by state GT strength, branching ratio (I) and partial half-life (in unit of $s$) for  $\beta^+$ decay of $^{27}$P employing the three different  pairing gap (TF, 3TF and 5TF) values.}
	\begin{tabular}{cccc|cccc|cccc}
		\hline
		\multicolumn{4}{c}{TF} & \multicolumn{4}{c}{3TF} &\multicolumn{4}{c}{5TF}\\
		\hline
		E$_x$ (MeV)& GT & I&t$^{par}_{\textcolor{red}{1/2}}$&E$_x$ (MeV)& GT & I&t$^{par}_{\textcolor{red}{1/2}}$&E$_x$ (MeV)& GT & I&t$^{par}_{\textcolor{red}{1/2}}$ \\
		\hline
		0.782  & 0.0093 & 16.59 & 4.10E+00 & 0.744  & 0.0285 & 93.68 & 1.14E+01 & 0.781  & 0.0057 & 14.29 & 6.66E+00 \\
		0.959  & 0.0005 & 0.93  & 7.33E+01 & 0.942  & 0.0040 & 5.64  & 1.89E+02 & 0.976  & 0.0003 & 0.83  & 1.14E+02 \\
		2.911  & 0.0363 & 42.11 & 1.62E+00 & 2.963  & 0.0002 & 0.16  & 6.64E+03 & 2.867  & 0.0217 & 33.18 & 2.87E+00 \\
		3.591  & 0.0394 & 17.25 & 3.95E+00 & 3.576  & 0.0104 & 0.46  & 2.33E+03 & 3.650  & 0.0221 & 12.46 & 7.63E+00 \\
		3.838  & 0.0785 & 15.43 & 4.41E+00 & 3.970  & 0.0114 & 0.07  & 1.59E+04 & 3.795  & 0.0427 & 10.48 & 9.07E+00 \\
		5.267  & 0.0000 & 0.00  & 7.01E+04 & 5.115  & 0.0001 & 0.00  & 1.52E+09 & 5.291  & 0.0001 & 0.01  & 2.05E+04 \\
		5.395  & 0.0064 & 0.14  & 4.98E+02 & 5.551  & 0.0001 & 0.00  & 1.17E+09 & 5.422  & 0.0040 & 0.27  & 3.52E+02 \\
		6.326  & 0.0134 & 0.28  & 2.45E+02 & 7.269  & 0.0004 & 0.00  & 2.56E+08 & 5.525  & 0.0028 & 0.19  & 5.03E+02 \\
		6.328  & 0.0137 & 0.29  & 2.39E+02 & 10.468 & 0.0148 & 0.00  & 1.05E+07 & 5.617  & 0.0001 & 0.01  & 2.02E+04 \\
		6.541  & 0.1270 & 2.05  & 3.32E+01 & 10.469 & 0.0104 & 0.00  & 1.49E+07 & 5.835  & 0.1603 & 7.92  & 1.20E+01 \\
		6.654  & 0.3064 & 4.31  & 1.58E+01 & 10.583 & 0.0292 & 0.00  & 7.08E+06 & 5.935  & 0.3651 & 16.28 & 5.84E+00 \\
		7.328  & 0.0020 & 0.01  & 6.00E+03 & 10.708 & 0.0166 & 0.00  & 1.80E+07 & 6.582  & 0.0022 & 0.05  & 2.00E+03 \\
		7.469  & 0.0094 & 0.04  & 1.61E+03 & 10.972 & 0.0698 & 0.00  & 1.30E+07 & 7.177  & 0.0140 & 0.14  & 6.94E+02 \\
		7.891  & 0.0110 & 0.02  & 2.80E+03 & 11.047 & 0.0019 & 0.00  & 7.43E+08 & 7.582  & 0.3880 & 2.04  & 4.66E+01 \\
		8.134  & 0.3679 & 0.51  & 1.33E+02 & 11.276 & 0.0380 & 0.00  & 1.09E+09 & 7.658  & 0.3070 & 1.43  & 6.67E+01 \\
		8.733  & 0.0206 & 0.01  & 9.53E+03 &   ---     &  ---      &  ---     &   ---       & 8.125  & 0.0313 & 0.06  & 1.54E+03 \\
		8.799  & 0.0119 & 0.00  & 1.97E+04 &   ---     &  ---      &  ---     &   ---       & 8.170  & 0.0004 & 0.00  & 1.21E+05 \\
		8.882  & 0.0001 & 0.00  & 3.41E+06 &   ---     &  ---      &  ---     &   ---       & 8.202  & 0.0143 & 0.02  & 3.93E+03 \\
		8.947  & 0.0027 & 0.00  & 1.35E+05 &   ---     &  ---      & ---      &   ---       & 8.346  & 0.0006 & 0.00  & 1.35E+05 \\
		9.115  & 0.2334 & 0.03  & 2.68E+03 &   ---     &  ---      &  ---     &   ---       & 8.522  & 0.2350 & 0.20  & 4.87E+02 \\
		9.512  & 0.5321 & 0.01  & 5.91E+03 &   ---     &  ---      &  ---     &   ---       & 8.962  & 0.5413 & 0.14  & 7.02E+02 \\
		9.937  & 0.0313 & 0.00  & 1.06E+06 &   ---     &  ---      &  ---     &   ---       & 9.300  & 0.0393 & 0.00  & 3.17E+04 \\
		10.074 & 0.0173 & 0.00  & 3.67E+06 &   ---     &  ---      &  ---     &   ---       & 9.385  & 0.0264 & 0.00  & 6.70E+04 \\
		10.240 & 0.1783 & 0.00  & 5.42E+05 &   ---     &   ---     &  ---     &   ---       & 9.694  & 0.1495 & 0.00  & 5.42E+04 \\
		10.804 & 0.0000 & 0.00  & 2.68E+10 &   ---     &   ---     &  ---     &   ---       & 10.176 & 0.0000 & 0.00  & 3.29E+09 \\
		11.006 & 0.0118 & 0.00  & 9.38E+07 &   ---     &  ---      &  ---     &   ---       & 10.385 & 0.0080 & 0.00  & 1.60E+07 \\
		11.026 & 0.0060 & 0.00  & 2.10E+08 &   ---     &  ---      &  ---     &   ---       & 10.399 & 0.0050 & 0.00  & 2.66E+07 \\
		11.138 & 0.0018 & 0.00  & 1.79E+09 &   ---     &  ---      &  ---     &   ---       & 10.542 & 0.0011 & 0.00  & 1.74E+08 \\
		\hline
	\end{tabular}
\end{table}

\begin{figure}
	\hfil
	\vspace*{-5cm}
\hspace*{-5cm}	\includegraphics*[width=1.3\paperwidth]{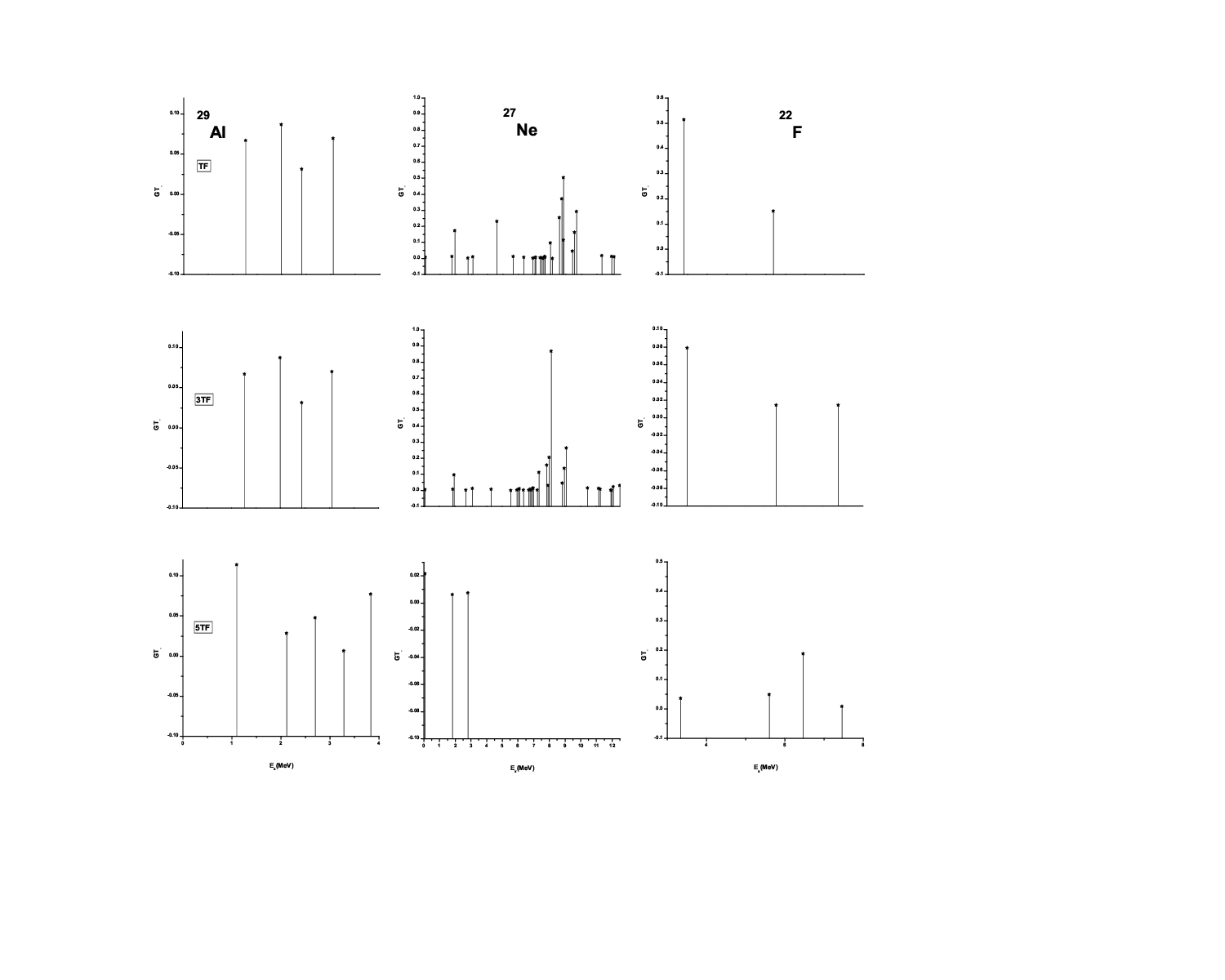}
\vspace*{-5cm}
\scriptsize \caption{Computed GT strength distributions for $^{29}$Al, $^{27}$Ne and $^{22}$F as a function of excitation energy  in daughter, employing the pn-QRPA model with different pairing gap (TF, 3TF and 5TF) values.}
\end{figure}\label{F1}
\begin{figure}\label{F2}
	\hfil
	\vspace*{-5cm}
	\hspace*{-5cm}	\includegraphics*[width=1.3\paperwidth]{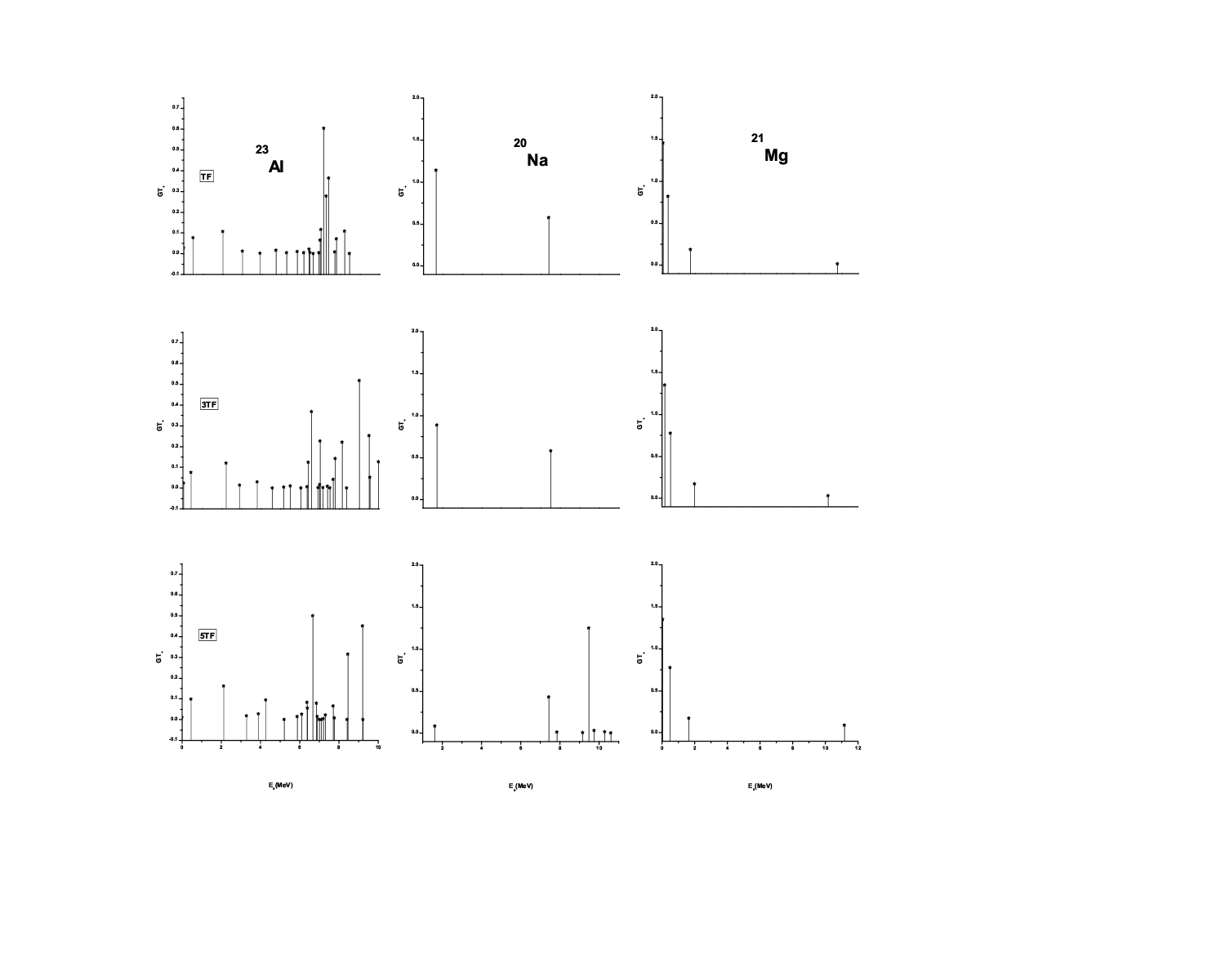}
	\vspace*{-5cm}
	\scriptsize \caption{Same as Fig.~1 but for $^{23}$Al, $^{20}$Na and $^{21}$Mg.}
\end{figure}
\newpage
\section{Summary and Conclusion}
We presented the calculation of GT strength distributions and $\beta$-decay half-lives for a total of 47 $sd$-shell nuclei in the mass range 20 $<$ A $<$ 30 employing the pn-QRPA model. Three different formulae were used to compute the pairing gaps between the nucleons in order to investigate their effect on the calculated GT strength distributions and half-lives. The GT strength distributions and centroids were altered, in few cases, substantially, with change in pairing gap values. The $\beta$-decay half-lives computed using the mass dependent formula (TF) were in overall better comparison with the measured data. For few cases, studied in this project, 3TF and 5TF formulae for pairing gaps reproduced the measured half-lives the best.

\clearpage

\bibliography{mybibfile}

\end{document}